\documentclass[prl,aps,preprint,groupedaddress,floats]{revtex4}

\usepackage{graphicx}
\usepackage{dcolumn}
\usepackage{bm}
\usepackage{amsmath}
\usepackage{color}
\usepackage{url}
\newcommand{\be}{\begin{equation}}
\newcommand{\ee}{\end{equation}}

\def\he4{$^4$He}
\def\hel3{$^3$He}

\begin{document}

\title{Transverse Quantum Superfluids}

\author{Anatoly Kuklov}
\affiliation{Department of Physics and Astronomy, The College of Staten Island, and the Graduate Center, CUNY, Staten Island, NY 10314, USA}

\author{Lode Pollet}
\affiliation{Arnold Sommerfeld Center for Theoretical Physics, Ludwig-Maximilians Universit\"at, Theresienstrasse 37, 80333 M\"unchen, Germany}
\affiliation{Munich Center for Quantum Science and Technology (MCQST), 80799 M\"unchen, Germany}

\author{Nikolay Prokof'ev}
\affiliation{Department of Physics,  University of Massachusetts, Amherst, MA 01003, USA}

\author{Boris Svistunov}
\affiliation{Department of Physics,  University of Massachusetts, Amherst, MA 01003, USA}

\begin{abstract}
Even when ideal solids are insulating, their states with crystallographic defects may have superfluid properties. It became clear recently that edge dislocations in \he4 featuring
a combination of microscopic quantum roughness and superfluidity of their cores may represent a new paradigmatic class of 
quasi-one-dimensional superfluids. The new state of matter, termed transverse quantum fluid (TQF), is found in a variety 
of physical setups. The key ingredient defining the class of TQF systems is infinite compressibility, which is responsible for all other unusual properties such as the quadratic spectrum (or even the absence) of normal modes, irrelevance of the Landau criterion, off-diagonal long-range order at $T = 0$, 
and the exponential dependence of the phase slip probability on the inverse flow velocity. From a conceptual point of view, the TQF state is a striking demonstration of the conditional character of many dogmas associated with superfluidity, including the necessity of elementary excitations, in general, and the ones obeying Landau criterion in particular.  
\end{abstract}

\maketitle

\section{INTRODUCTION}

One and a half decade ago experiments discovered, quite unexpectedly, the mysterious effect of the anomalous isochoric compressibility (a.k.a. syringe effect) in \he4. 
Even though first-principle simulations almost immediately provided a successful interpretation in terms of the superclimb ({\it i.e.}, the climb motion of edge dislocations supported by superfluidity in their cores), a fully satisfactory explanation of all facets remained elusive. It became clear only recently that a microscopically quantum-rough edge dislocation with a superfluid core is just one paradigmatic example of a broad class of novel quasi-one-dimensional (1D) superfluids. The (almost detective) story and the broader context of the experimental discovery, as well as the history of its theoretical interpretation, are both intriguing and instructive. An extended synopsis, sufficient as an introduction to the field for newcomers, follows in the subsections below. A more comprehensive review is provided in Sections 2-4.

\subsection{Experiment never lies...}

Back in 2004, E. Kim and  M. H. W. Chan reported their exciting experimental finding: 
Below a temperature of $\approx 0.2$ K, the period of a torsion oscillator filled with solid \he4 demonstrates an abrupt drop \cite{Kim_Chan_2004}.
This was originally attributed to mass decoupling, the smoking gun of the phenomenon of supersolidity, which is the ability of a solid to support supertransport of its own atoms. However, since it was known that the ground state of solid \he4 is a {\it commensurate crystal} \cite{commens}, a perfect-crystal supersolid interpretation was rigorously ruled out \cite{PS2005}.
Indeed, a compressible supersolid is only possible in the presence of zero-point vacancies (or interstitials, or both), thus proving that there is no fundamental alternative to 
E. Gross's \cite{Gross57, Gross58} classical-density-wave picture, which takes the form of the Andreev-Lifshitz-Chester \cite{Andreev_Lifshitz,Chester} zero-point vacancy scenario
in the quantum limit. In the absence of particle-hole symmetry or extreme fine-tuning a supersolid crystal has to be incommensurate.
First-principle path-integral simulations by Ceperley and Bernu \cite{Ceperley_Bernu} also cooled down the original excitement  by explicitly demonstrating that a perfect \he4  crystal has no 
off-diagonal long-range order (ODLRO).

The challenge of properly interpreting Kim and Chan's discovery resulted in a number of experimental and theoretical activities pursuing two radically different---apparently unrelated---hypotheses: (i) defect-induced supersolidity and (ii) a low-temperature elastic anomaly potentially pointing to a novel quantum phenomenon. A detailed account of various aspects of these activities can be found in review papers \cite{BP2012,Hallock2019,Beamish_Balibar2020,Yukalov2020,Chan2021,Fil_Shevchenko2022}. In what follows, we mention the crucial findings revealing the phenomena directly or circumstantially related to the topic of transverse quantum fluids: defect-induced supersolidity and microscopic quantum roughness of 1D defects.

\subsection{Defect-induced supersolidity}

While not being widely known/appreciated, putting aside direct experimental and/or {\it ab initio}  theoretical interests, the idea of defect-induced supersolidity in  \he4 crystal 
had been formulated long ago by S. Shevchenko  who pointed to a hypothetical possibility for dislocations to have superfluid cores and envisioned a superfluid network of such dislocations \cite{Shevchenko1987}. 
First-principle simulations of various defects in a \he4 crystal performed by our collaboration using continuous-space Worm algorithm \cite{Worm2006} revealed superfluidity in certain grain boundaries \cite{GB} and screw dislocations \cite{screw} (see Section~\ref{sec:defects}). A metastable superglass phase was also observed \cite{superglass}. Later---motivated by the results of the UMass Sandwich experiment (see below)---superfluidity in cores of certain edge dislocations was also found \cite{sclimb}.

\subsection{Quantum roughness and interplay with superfluidity}

The original motivation behind the study of the interface
between two checkerboard solid domains on a lattice \cite{Burovski2005} was to find a proof-of-principle numeric evidence for the scenario of defect-induced supersolidity. Unexpectedly, simulations discovered a far more delicate phenomenon---an emergent conspiracy between the superfluidity and interface roughness: the superfluid-to-insulator transition occurs simultaneously with the rough-to-smooth transition in terms of the coarse-grained interface shape \cite{Burovski2005,Soyler2007}. The theoretical explanation is based on the conjecture that criticality is driven by proliferation of spinon excitations 
that simultaneously carry two emergent quantum numbers: fractional particle charge of $\pm 1/2$ and a quantum of interface displacement---an elementary kink \cite{Burovski2005}.

In the context of present review, this result points to the possibility of nontrivial interplay between two apparently unrelated types of motions in a quasi-1D superfluid with transverse degrees of freedom: supertransport and quantum fluctuations of the shape. The character of this interplay differs fundamentally between the checkerboard interface 
and the superclimbing edge dislocation \cite{sclimb} indicating 
the crucial role played by infinite compressibility in the latter 
system. 

\subsection{Quantum plasticity} 

In 2007, Beamish and Day reported an observation of a shear modulus anomaly in solid \he4~\cite{Day_Beamish_2007}.  This result took the community by surprise and led to the questioning of the supersolid interpretation of the experiment by Kim and Chan. The characteristic form of the shear response (with a reversible reduction of up to 90\% in the shear modulus) was found to follow closely the torsion oscillator period data. It should be noted that the edge dislocations responsible for such a reduction of the shear modulus are different from the superclimbing ones by the orientation of their Burgers vectors \cite{Hull}---in the basal plane and along the C-axis, respectively. Furthermore, while the former are insulating and can only glide at low $T$, the latter have superfluid core and can perform superclimb.
Further investigations of elastic properties of imperfect \he4 crystals converged into the notion of giant quantum plasticity \cite{Rojas2010,Haziot}. Besides explaining the findings of Reference~\cite{Kim_Chan_2004}  [for a comprehensive discussion, see Reference~\cite{Chan2021}], the  phenomenon reveals fundamentally new physics, which is arguably more exciting than conventional supersolidity. To persist down to extremely low temperature, as suggested by experimental observations \cite{Haziot},  the effect of giant quantum plasticity has to be supported by microscopically quantum-rough basal edge dislocations; 
otherwise dislocations would be trapped by Peierls barriers at sufficiently low temperatures and the effect would disappear. The superclimbing modes and quantum renormalization of the shear modulus emerge as twin phenomena both requiring---and thus revealing---microscopic quantum roughness of edge dislocations.

\subsection{UMass Sandwich, supertransport, syringe effect, superclimb}\label{sec:sandwich}

\begin{figure}[htbp]
\centering
\includegraphics[width=1.0 \columnwidth]{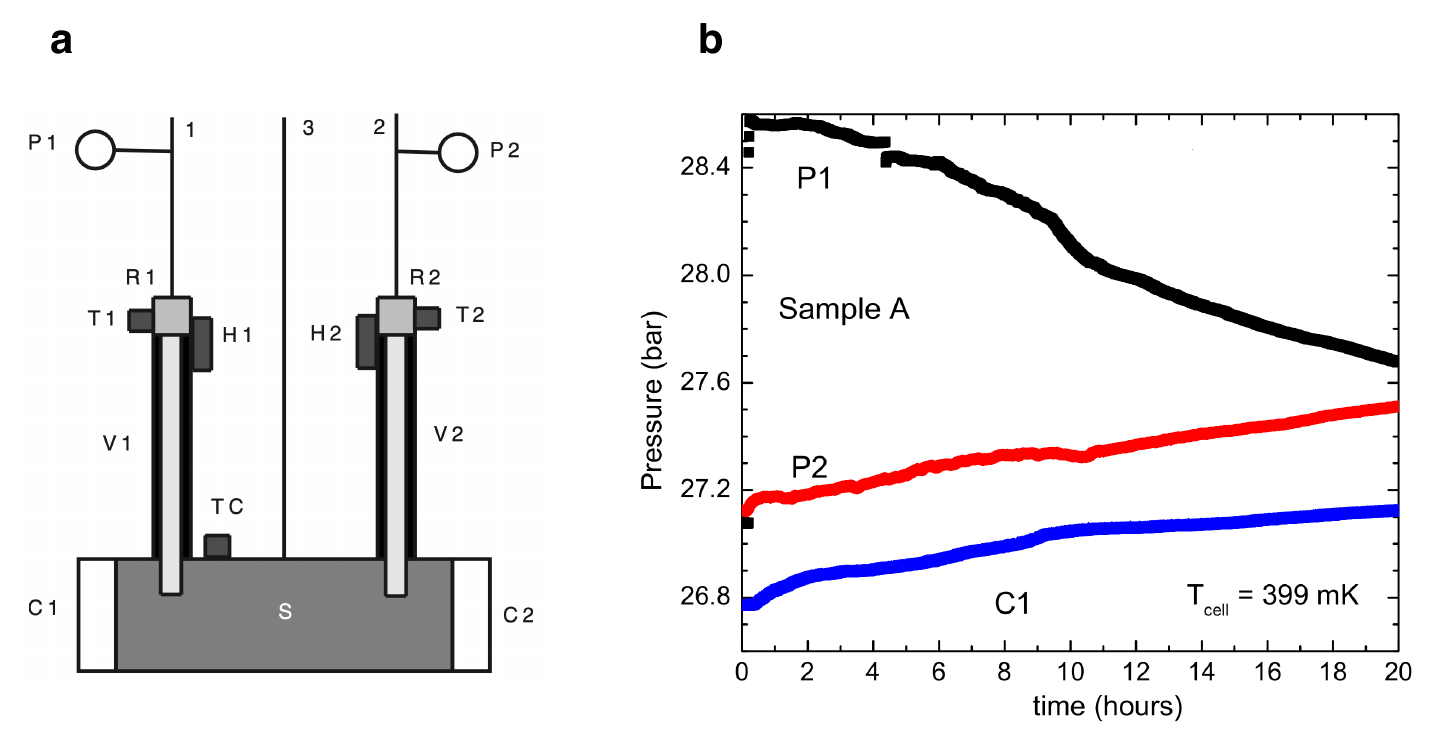}
\caption{UMass Sandwich of Ray and Hallock: Schematic diagram of the experimental cell and the reported data. (a) Three filling lines lead to the cell. Two of them go to liquid reservoirs R1 and R2 above the Vycor rods V1 and V2, while the third one goes directly to the solid chamber, S. Two capacitance pressure gauges, C1 and C2, sit on either side of the cell for in situ pressure measurements. Each reservoir has a heater, H1 and H2, which prevents the liquid in it from freezing, and the reservoir temperatures are read by carbon resistance thermometers T1 and T2. The cell temperature is recorded by a third carbon resistance thermometer, TC. The cell thermometer reading, denoted as TC, provides the temperature of the sample, T. (b) Response of the apparatus to a pressure step applied to line 1 for sample A, grown at 26.8 bar. The shift in P1 just after 4 h is an artifact from the fill of a nitrogen trap. The regulator, which controls the pressure that feeds line 1, was shut off at 6 h. The pressure increase in line 2 was essentially linear in time, especially for the final 10 h. Panels {\it a} and {\it b} with captions are adapted with permissions from References \cite{Hallock2009} and \cite{Hallock2008}, respectively.}
\label{Sandwich}
\end{figure}

In an attempt to present direct evidence for the ability of \he4 crystals to support superflow of its own atoms,
Hallock and Ray developed an ingenious experimental setup (shown in {\bf Figure \ref{Sandwich}a}) called the UMass Sandwich, allowing one to study DC transport, if any, through solid \he4  \cite{Hallock2008,Hallock2009}. The idea behind the UMass Sandwich is to connect two liquid superfluid reservoirs, the source and the drain, with a solid \he4 sample via Vycor rods playing the role of leads. The crucial trick is the temperature gradient  along the rods allowing one to keep two different phases, the solid and the liquid, at one and the same chemical potential. 

The results proved to be both exciting and puzzling. The direct evidence of supertransport (see {\bf Figure \ref{Sandwich}b}) was obtained in the form of the characteristic supercritical regime---linear dependence of the pressure difference on time. A striking discovery,  creating an impression of an ``experimental bug," was the effect of the giant isochoric compressibility, a.k.a.  syringe effect: In all samples supporting superflow, the pressure was also changing in the solid bulk, i.e. the pressure/density of the constant-volume solid was responding to small changes of the applied chemical potential; see the curve C1 in {\bf Figure \ref{Sandwich}b}. The unique experimental protocol enabled by the UMass Sandwich setup demonstrated that the solid response was no different from that of a liquid.  

The only way an ideal crystal can change its pressure/density 
at constant volume and temperature is by changing its interatomic distance $a$ by adding/removing complete atomic planes. In an insulating 
solid, such a mechanism is inhibited at low temperature. The only existing scenario explaining the syringe effect is based on superclimb of edge dislocations \cite{sclimb}. In contrast to the conventional climb assisted by pipe diffusion of thermally activated vacancies along the dislocation core \cite{Lothe,Hull} (viable only at high temperature), or (less conventional but still ``normal" in the context of our discussion) climb of edge dislocations in bacterial cell walls \cite{Amir_Nelson}, the low-temperature climb of edge dislocations in \he4 must be assisted by the superflow along the core. The syringe effect persisting down to low temperatures when thermal activation of vacancies is no longer possible, as well as first-principle simulations of edge dislocations demonstrating superclimb \cite{sclimb}, provide strong support to the minimalistic unified scenario behind all observed phenomena based on the microscopically quantum-rough edge dislocations (with their Burgers vector along the C-axis) with superfluid cores.

The syringe effect naturally explains results of ``inverted" setups when matter gets squeezed out from one solid \he4 reservoir (the source) into the other solid reservoir  (the drain)  through either Vycor or solid leads. Both versions of the inverted syringe setup generating the compression-driven DC superflow were implemented by Cheng, Beamish, Balibar, and their collaborators \cite{Cheng2015,Cheng2016}.

\subsection{Critical flux: mysterious temperature dependence} 
\begin{figure}[htbp]
\includegraphics[width=1.0 \columnwidth]{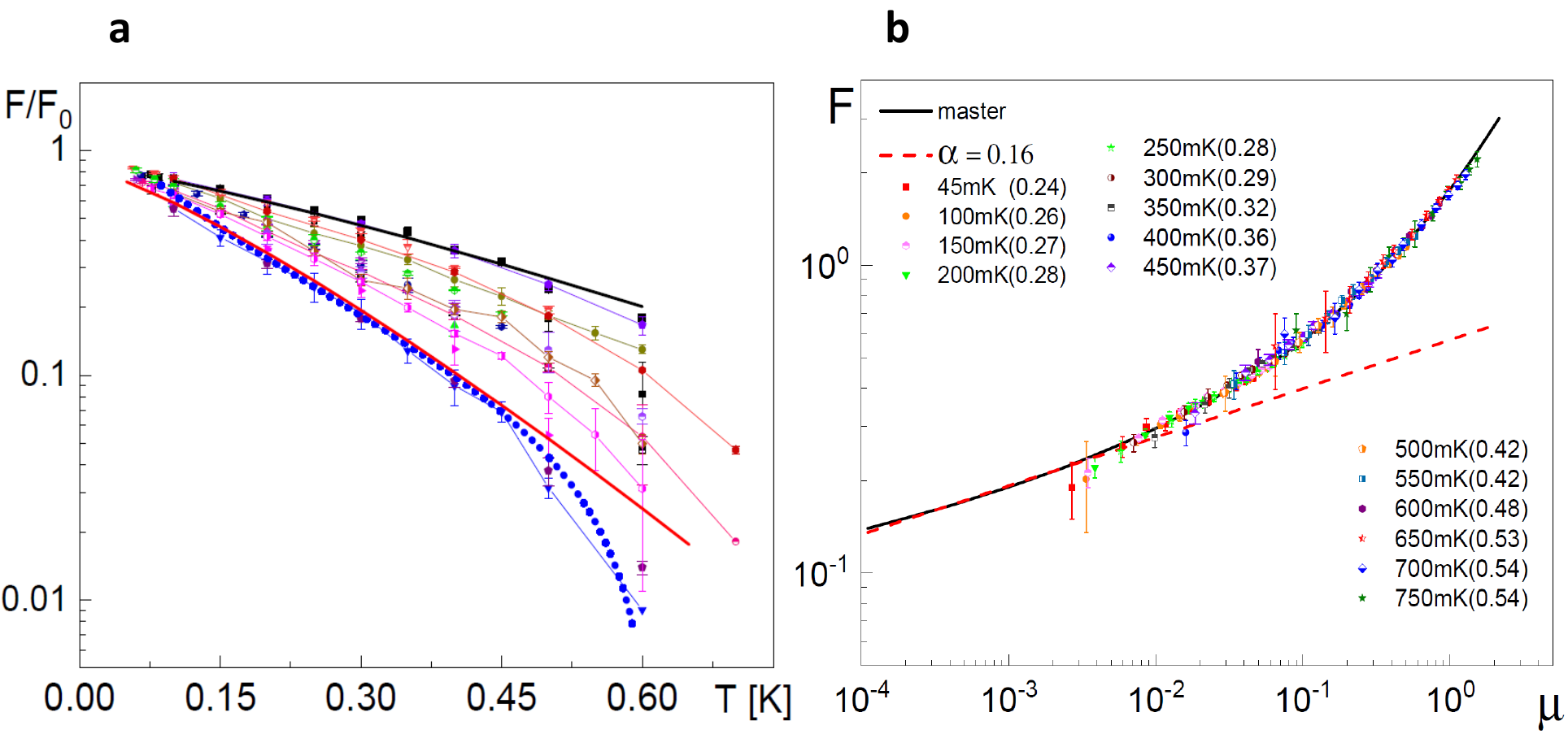}	
\caption{Experimental data for the critical flux $F$ versus temperature, $T$, and the chemical potential bias, $\mu$.  (a) $F(T)$ normalized by $F_0 = F(T\approx 0)$ for differently grown \he4 samples. Data points are from  References~\cite{Moses,Moses2019} (lines are guide to the eye). The dotted line represents the master curve \cite{Hallock2014} fitting the data for multiple samples.  
All the data for $T<0.5$ K are consistent with the stretched exponential law $\exp[-(T/T_{\alpha})^{\alpha}]$, $\alpha =$ 1--1.3,
predicted by the model of Reference~\cite{Kuklov2022}: The thick solid lines are fits with $\alpha=5/4$ and $T_{5/4}\approx 0.20K$, $0.45K$ for the
lowest and the highest data sets, respectively, \cite{Moses,Moses2019}.
(b) The solid line is the master I-V curve $v\exp(-1/v)=\mu/v_c$ based on the probability of quantum phase slips, Equation \ref{P_exp}, where the flow velocity $v$ is expressed in units of $v_c$ and $\mu $ is scaled by a constant factor \cite{Radzihovsky2023}.  Symbols are experimental data from Reference~\cite{Moses2019} for different $T$ and interpreted using the sub-Ohmic dependence $v\propto \mu^\alpha$.  The corresponding values of $T$ and $\alpha$ (in parenthesis) are shown in the legend. Data points for a given $T$ in the log-log scale are shifted to achieve the best fit with the master curve without changing the slope. The dashed line corresponds to the power law  with $\alpha =0.16$ shown for comparison. While the master curve is different from the LL behavior $v\sim \mu^\alpha$, parts of the master curve (about two orders of magnitude in the bias) can be fit by the sub-Ohmic dependence with $\alpha$ increasing from $\alpha\approx 0.2$ at low $T$ to $\alpha\approx 0.5$ at large $T$. Panels {\it a} and {\it b} are adapted with permissions from References \cite{Kuklov2022} and \cite{Radzihovsky2023}, respectively.}
\label{fig:anomalous_critical_flux}
\end{figure} 

Given that  dislocation cores are quasi-1D objects, it was natural to describe their superfluid properties within the Luttinger liquid (LL) paradigm when at zero
temperature the I-V curve is sub-Ohmic at small bias. 
This expectation  resulted in a widely shared point of view that supertransport of \he4 atoms is through the network 
of dislocations behaving as bosonic LLs \cite{Hallock2019,Chan2021}. However, at low $T$, the initial part of the otherwise temperature-independent I-V curve in LL
is Ohmic and characterized by high conductivity diverging in the $T\to 0$ limit as a power law. In contrast, experiments consistently observed a different unique temperature dependence of the critical flux $F$ shown in {\bf Figure~\ref{fig:anomalous_critical_flux}}, forcing one to speculate about yet unknown exotic properties of dislocation networks; cf.~\cite{KPS2011}. 

\subsection{Transverse quantum fluid}

It was realized only recently \cite{Kuklov2022}  that the mysterious temperature dependence of the critical flux perfectly fits into the new paradigm of quasi-1D superfluid, the so-called transverse quantum fluid (TQF) \cite{Radzihovsky2023,Kuklov2024a,Kuklov2024b}. The key feature defining the class of TQF systems is their infinite compressibility, which in the case of the edge dislocation is associated with the superclimbing degree of freedom: When particles are added to the core, the latter  climbs in the transverse direction without energy penalty.  
Unusual properties of TQFs include: quadratic dispersion (or even absence) of normal modes, irrelevance of Landau criterion, ODLRO at $T = 0$, and exponential dependence of the phase slip probability on the inverse flow velocity allowing TQF to support finite supercurrents. It turns out that superclimbing edge dislocation is not the only system that can be characterized as TQF: there exists a broad class of quasi-1D superfluids featuring similar properties \cite{Kuklov2024a,Kuklov2024b}. Examples include a superfluid edge of a self-bound droplet of hard-core bosons on a two-dimensional (2D) lattice, a Bloch domain wall in an easy-axis ferromagnet, a phase separated state of two-component bosonic Mott insulators with the boundary in the counter-superfluid phase (or in a phase of two-component superfluid) on a 2D lattice, and edges of incomplete surface layers (on substrates or crystalline facets). 
From a conceptual point of view, the TQF state is a striking demonstration of the conditional character of many dogmas associated  with superfluidity, including the necessity of elementary excitations, in general, and the ones obeying the Landau criterion in particular.  

The explanation of the temperature dependence of the critical flux \cite{Kuklov2022,Radzihovsky2023} involves two ingredients: (i) thermal fluctuations of the edge dislocation shape 
inducing large, correlated, and asymmetric stress fields acting on the superfluid core and (ii) exponential sensitivity of local superfluid properties on the stress field fluctuations with
bottlenecks defining the critical flux. 
 
 \subsection{Outline} 
 
The rest of this review is organized as follows.  In Section ~\ref{sec:2}, we discuss  {\it ab initio} evidence for the supertransport and syringe effects in imperfect \he4 crystal. 
We show how the latter is naturally explained in terms of the superclimb of microscopically quantum-rough edge dislocations 
with superfluid cores, and present arguments leading to the formulation of the effective field theory for superclimbing modes.
In Section ~\ref{sec:3}, we formulate the effective field theory
and demonstrate that it represents a new TQF paradigm for quasi-1D superfluidity. We reveal basic properties of TQF 
and show that similar physics can be observed in a number 
of other settings, which, despite being less exotic than superclimbing edge dislocations, were overlooked in the past. 
We then show that superclimb-based  TQFs are closely related to other quasi-1D superfluids named incoherent transverse quantum fluids (iTQF). The iTQF systems (some of which being well-known) feature gapless transverse particle reservoirs guaranteeing infinite compressibility of the effective 1D field theory and thus unusual properties characteristic of TQFs. However gapless particle reservoirs also lead to a fundamental difference: strong dissipation eliminates normal modes in the iTQF superfluid and the evolution of the disturbed field of superfluid velocity is incoherent (diffusive). The hallmark properties shared by both TQFs and iTQFs are the ODLRO (Bose-Einstein-condensed) ground state and power-law binding of instantons allowing both systems to support finite supercurrents (at low enough temperature).
In Section ~\ref{sec:4}, we go back to the supertransport in solid \he4. Using the theory of TQF instantons we show how the 
temperature dependence of the critical flux fits into the TQF paradigm.  
We conclude with summarizing the results and discussing further experimental and theoretical steps towards full understanding of 
TQF and iTQF physics.

\section{SUPERCLIMBING EDGE DISLOCATIONS IN HE-4}
 \label{sec:2}
 
In this section, we review the physics of superclimbing edge dislocations in \he4. We start by putting them in the wider context of all defects known to have the capacity of suporting superfluidity in \he4 (Section~\ref{sec:defects}). This brief summary is followed by the theoretical concept of superclimb (Section~\ref{sec:syringe_th}).

\subsection{Defect-induced superfluidity in solid \he4: numerical evidence} \label{sec:defects}

After a short period of controversial numerical results all methods
converged on the conclusion that the perfect crystalline phase of (hcp) solid \he4 is a commensurate insulator~\cite{Ceperley_Bernu,superglass,nogo3}. Therefore, 
an explanation of the experimental superflow-through-solid (STS) observation requires crystalline defects and a proof that 
they can support supercurrents. 

Zero-dimensional defects such as vacancies have a relatively 
large energy gap~\cite{fate2006} (interstitials have a much larger gap and effective mass than vacancies and are, therefore, of secondary importance) implying that their thermal population 
at $T<0.5$ K is negligible. Even non-equilibrium 
vacancies cannot support the Andreev-Lifshitz-Chester scenario~\cite{Andreev_Lifshitz} because vacancies in hcp \he4
attract each other~\cite{fate2006,Rossi2008, Rossi2009}. In the thermodynamic limit, the crystal would purge itself of such defects through phase separation or formation of dislocation loops. Similar behavior was also seen in two dimensions~\cite{Rossi2010,Rossi2012}.

Two-dimensional defects such as grain boundaries were found to be generically stable and superfluid with a transition temperature of the order of $\sim 0.5$ K ~\cite{GB}. 
One-dimensional defects, in particular dislocations, were identified 
as the most important ones because of their ability to form robust networks. The screw dislocation with the axis oriented along the $c-$axis features a superfluid core of diameter of about $\sim 6$ \AA \, near the melting line~\cite{screw}. 
Edge dislocations with in-plane Burgers vectors are 
insulating (as are grain boundaries with low tilt angle).  
However, edge dislocations with Burgers vector along the $c-$axis and in-plane orientation behave differently: they split into two partials with a large fcc fault in between (estimated to be $\sim 300$ \AA~wide). Each partial has a superfluid core and responds 
to chemical potential changes, i.e. it demonstrates superclimb~\cite{sclimb}. 
The necessary condition for superclimb is microscopic quantum roughness of the core; if the core is wide enough the Peierls 
barrier can be neglected all the way to the thermal 
wavelength of superclimbing modes, {\it i.e.}, the dislocation behaves akin to a free string. 

A physical picture explaining under what conditions 
crystallographic defects are expected to have superfluid properties 
was put forward in Reference~\cite{stress2008} and supported
by all studies mentioned above: the strain field caused by the defect reduces the gap for vacancy formation. If the gap closes, vacancies condense on the defect, and cause superfluidity.  

When \he4 is confined to nano-cylinders, it shows a shelled structure reminiscent of smectic-A liquid crystal containing Frank’s disclination. Superflow along the melted disclination core is stable at low pressure and remains extremely metastable at pressures exceeding the spinodal for overpressurized bulk superfluid \he4. Thus, in nanoporous materials, disclinations are the likely candidates for controlling the superflow properties of \he4~\cite{disclination}. We note that previous studies of \he4 also argued in favor of liquid-like behavior induced by confined geometry or boundary ~\cite{Khairallah2005,DelMaestro2011,Boninsegni_nano2015}.  

\subsection{Superclimb: theoretical concept} \label{sec:syringe_th}

An edge dislocation can perform two types of motion: glide and climb \cite{Hull}. Glide is a transverse motion that conserves the total amount of matter. Climb, by contrast, requires matter to be added or removed in order to grow or shrink an incomplete atomic plane of which the dislocation is the boundary. In classical materials, matter is carried out by vacancies (or interstitials) either from the bulk or along the dislocation core (pipe diffusion), leading to classical plasticity. This activation process strongly depends on temperature and in solid \he4 stops at $T<0.5$ K. The twin observations of giant plasticity and isochoric compressibility at lower temperature force one to consider quantum-mechanical effects. Assuming for a moment that a dislocation is akin to a freely oscillating string, it corresponds, at the microscopic level, to a quantum rough object, that is best thought of as a liquid of kinks and jogs.  Quantum glide corresponds to barrier-free motion of kinks (it does not require superfluidity) and results in the reversible ``giant" plasticity. In the quantum climb motion 
the pipe diffusion is replaced by the matter flow along the superfluid core, hence the name superclimb \cite{sclimb}. Since matter leads to the core displacement $h$ in the climb direction, 
$h$ is a canonically conjugate variable to the superfluid phase $\phi$. Furthermore, translation symmetry dictates that the 
free-string energy dependence on $h$ is of the  Granato-L\"ucke form~\cite{GranatoLuecke}, $\sim (\partial_x h)^2$
(cf. Section~\ref{sec:3} below)---as opposed to $\sim h^2$ for 
an atomic-scale smooth dislocation or LL. In response to the chemical potential bias $\delta \mu$, the climb displacement 
is then diverging with the dislocation length $L$ as 
$ h\propto L^2 \delta \mu$. Since the number of added/removed atoms to the atomic layer is $\delta N \propto h$, the 1D compressibility 
is diverging in the $L\to \infty$ limit as $\propto L^2$, explaining the syringe effect.   

In all known naturally occurring materials except for \he4 the Peierls potential is large and inhibits the proliferation and 
motion of kinks and jogs. An important figure of merit 
is the ratio between the energy of the interatomic potential to the zero-point energy. The former is typically 100 times larger than the latter, but for \he4 (with the interatomic potential of about $\sim 10$ K) this ratio is of the order of unity. [Vacancy tunnelling time in \he4 is estimated as $\sim 0.01$  ns.] It is also important that 
for cores with large diameter $d$ the Peierls barrier is suppressed exponentially in $(d/a)^2$ and gets irrelevant for all practical purposes very quickly (cf. Section~\ref{sec:3.2} below). 
Binding of \hel3 impurities to dislocations
\cite{Syshchenko2010,Corboz,Haziot,Haziot2013b,Kuklov_2014_prb}
may result in their pinning to the lattice, but this effect is 
sample preparation/purification dependent and transient on long time scales \cite{Moses2020}.

\section{TRANSVERSE QUANTUM FLUID}
\label{sec:3}

\subsection{Effective theory and paradigmatic models}
\label{sec:3.1}

 The minimal model capturing the physics of the superclimb includes two canonically congjugate fields, $\phi(x,\tau)$ and $h(x,\tau)$, where $x$ and $\tau$ stand for the coordinate along the core and (imaginary) time, respectively. At this point it is worth comparing TQF with LL \cite{Haldane} where  the energy functional depends on 1D density $n$ as $\sim n^2$ and the system has finite compressibility. In TQF the role of the density is taken by $h$, which accounts for how much matter was added to (or removed from) the atomic plane edge. Translation invariance in a perfect crystal prohibits an energy contribution $\sim h^2$, while the linear term $\sim h$ can be compensated by the chemical potential $\mu$. Accordingly, the TQF compressibility is infinite as long as there are no pinning centers violating translation invariance. 
 As will be discussed below, this feature determines 
 unique TQF properties \cite{Kuklov2024a}: the parabolic excitation spectrum, ODLRO at $T=0$, and stability of  supercurrents at low flow velocity $v_0$. These should be contrasted with the linear excitation spectrum, algebraic off-diagonal order, and phase-slips-dominated dynamics of DC currents in LL. 

The minimal Hamiltonian,  $H[\phi,h] = \int {\cal H} \, dx $, describing the low-energy physics of TQF is defined
by \cite{Radzihovsky2023}
\be {\cal H} =
{\chi \over 2} (\partial_x h)^2 + {n_s\over 2} (v_0 + \partial_x \phi )^2 \, ,
\label{H}
\ee 
where the first term is the energy of the dislocation core deformation with $\chi$ determined by the shear modulus of the crystal, and the second term accounts for the kinetic energy
of the superflow along the core, with $n_s$ standing for the superfluid stiffness.  

\begin{figure}[htbp]
\centering
\includegraphics[width=0.95 \columnwidth]{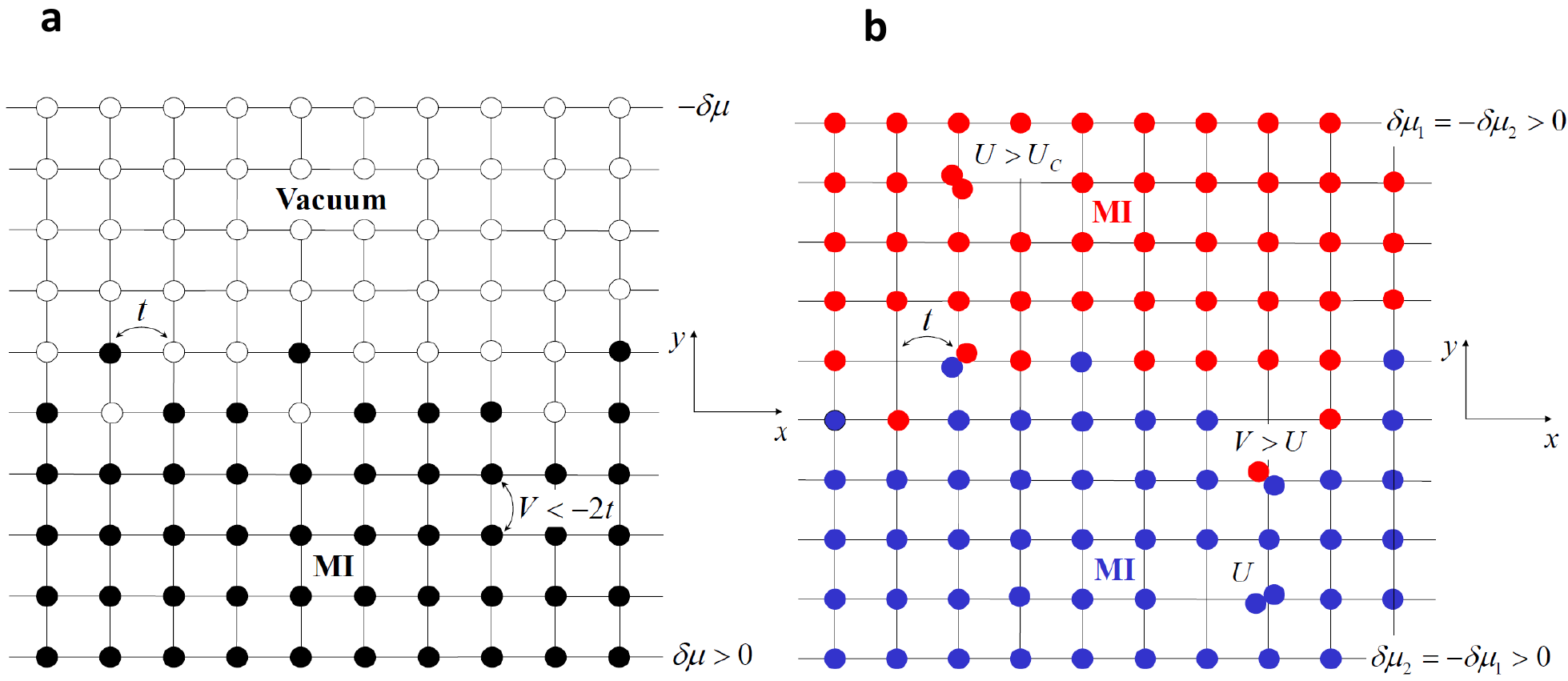}
\caption{(a) Self-bound system of hard-core bosons (solid circles) 
on the square lattice (open circles) with the phase boundary supporting the TQF state and performing superclimb in the direction transverse to the superflow. 
To orient the edge along the $x$-direction, pinning 
potentials $\pm \delta \mu$ of opposite sign are added at the lattice edges in the $y$-direction.
(b) Similarly designed and pinned two-component bosonic system in the phase separate state  (solid circles of different colors) at unit total filling. 
}
\label{fig:SCF}
\end{figure}

Among all possible TQF realizations the simplest one is arguably the 2D lattice containing hard core bosons ({\bf Figure~\ref{fig:SCF}a}). The Hamiltonian  consists of the nearest-neighbor
hopping and interaction terms on the square lattice:
\begin{equation}
H_{hc} = -t \sum_{\langle {i,j} \rangle } b_{j}^{\dagger} b_{i}^{\,}
        + V  \sum_{\langle {i,j} \rangle } n_{j} n_{i}    \;,
\label{Hhc}
\end{equation}
with the constraint on the occupation numbers, $n_{i}\le 1$ (here $b_{i}$ is the bosonic annihilation operator on site ${i}$).
In what follows, we use the hopping amplitude, $t$, and the lattice constant as units of energy and length, respectively.
At $V=-2t$ particles gain as much energy, $-4t$, from attractive
interactions in the MI state with $n_i=1$ as they get from delocalization in an empty
lattice, or vacuum state, with $n_{i} \to 0$. By decreasing $V$ below $-2t$, we ensure a
MI-vacuum phase-separated state with the width of the interface diverging as $V \to -2t$, which guarantees that the interface is in the superfluid phase.

The model \ref{Hhc} can be re-written identically as the spin-$1/2$ ferromagnetic $XYZ$-model \cite{Giamarchi}
\be
H_{XYZ}=\sum_{\langle {i,j} \rangle }[J_x S^x_i S^x_j+ J_y S^y_i S^y_j + J_z S^z_i S^z_j  ], 
\label{XYZ}
\ee
where $S^x_i, S^y_i, S^z_i$ are  spin-$1\over 2$ operators related to the bosonic operators as
$b^\dagger_i =S^x_i + iS^y_i, b_i =S^x_i -iS^y_i, S^z_i=b^\dagger_ib_i -{1 \over 2}  $, and 
$J_x=J_y=-2t$ and $J_z=V$. Thus, the phase separated
state of the model \ref{Hhc} corresponds to two neighboring domains with opposite orientations of spins, with the wide domain wall featuring the easy-plane polarization. 

The TQF physics is also expected to occur in the 2D lattice 
occupied by two species of repulsive bosons 
(labeled by index $\alpha=1,2$) at the total 
integer filling and in the regime of the phase separation between the components  ({\bf Figure~\ref{fig:SCF}b}). If the bulk of each spatially separated component is in the MI state, there is a regime when the boundary between the two species supports supercurrents with
counter-propagating flows of the components along the boundary,
{\it i.e.},  it is an edge in the supercounterfluid state \cite{SCF}. 
The corresponding microscopic Hamiltonian is given by
\be
H\, =\,  - t \!\!\!\! \sum_{\alpha=1,2; \langle i,j\rangle} \!\!\!\!  b^\dagger_{\alpha, i} b_{\alpha, j}  
\, + \,  \sum_i\left[{U\over 2}(n^2_{1,i} +n^2_{2,i}) + V n_{1,i}n_{2,i}\right] ,   
\label{SCF}
\ee
where $b_{\alpha,i}$ are bosonic annihilation operators, and $n_{\alpha,i} =b^\dagger_{\alpha,i}b_{\alpha,i}$. 
We consider the symmetric case when all parameters for different types of bosons are the same. The onsite interaction constants $U>0, V>0$ are chosen in such a way that $U$ exceeds the critical value $U_c$ for MI phase in the single-component system, and $V>U$ ensures phase separation when the two components are mixed. [If $V<U$ and $U>U_c$, the miscible state is in the supercounterfluid phase at low temperature \cite{SCF}.] 
The width of the supercounterfluid boundary between the two insulators is controlled by the proximity of $V$ to $U$  (at large enough $V_c$ the counter-transport ceases to exist). If the boundary is thick enough, the Peielrs barrier restricting its transverse motion can be neglected and we obtain an edge in the TQF regime. 

Another interesting system which may enter the TQF regime 
in its boundary is an incomplete layer of bosonic atoms on a 2D substrate or a facet of its own crystal. The description of this system is very similar to the hard-core attractive bosons discussed above. If the width of the incomplete layer edge
(often called a ``step") is wide in terms of the interatomic spacing, it can be in the TQF phase. In this regime, 
a droplet of incomplete layer becomes extremely mobile by transporting mass along the edge---an effect that may be revealed experimentally.

\subsection{Superclimbing modes and signature correlations}
\label{sec:3.2}

Perhaps the most striking feature of TQF is the direct
correspondence between the superfluid phase and geometrical shape fluctuations. One can literally measure the superfluid stiffness by ``looking" at  density snapshots.
This result immediately follows from the fact that
phase and density---and thus the ``vertical" position $h$ of the TQF edge---are canonically conjugate variables. 
By integrating out one of the canonically conjugate fields we obtain two equivalent actions each suitable for straightforward computation of the remaining field properties. Starting from 
\be
S[h,\phi]\, =\,    \int (i h \, \partial_\tau \phi \, +\,  {\cal H}) \, dx d\tau  
\label{Shphi}
\ee
with ${\cal H}$ given by Equation~\ref{H} (with $v_0$ set to zero), we readily obtain the height, $S_{h}[h]$, and phase, $S_{\phi}[\phi]$, actions from Gaussian integrals.  In the Fourier representation, we have
\be
S_h = { 1 \over 2}\sum_{\omega, k}   \left[n_s^{-1}\, {\omega^2 \over k^2}
  \, + \, \chi k^2 \right] |h_{\omega,k}|^2 \, ,
\label{S_eta}
\ee
\be
S_\phi = { 1 \over 2}\sum_{\omega, k}  \left[\chi^{-1}\, {\omega^2 \over k^2}
  \, + \, n_s k^2 \right] |\phi_{\omega,k}|^2 \, .
\label{S_phi}
\ee
As we already mentioned in Section~\ref{sec:3.1}, the superclimbing modes described by (any of) these actions have a quadratic dispersion \cite{sclimb}
\be
\omega_k = Dk^2 \, , \qquad D=\sqrt{n_s \chi} \, 
\label{omega_k}
\ee
and involve two ``orthogonal" types of quasi-one-dimensional motion
along and perpendicular to the edge: oscillations of the superflow and the geometric shape, respectively.

The simplest correlation functions revealing universal superclimbing fluctuations are
\be
K(x,\tau) \, =\, {1\over 2}\,  \langle \, [\, h(x,\tau)-h(0,0)\, ]^2 \, \rangle
\label{hight_corr}
\ee
and
\be
F(x,\tau) \, =\, {1\over 2}\,  \langle \, [\, \phi(x,\tau)-\phi(0,0)\, ]^2 \, \rangle \, .
\label{phase_corr}
\ee
As is evident from Equations~\ref{S_eta} and \ref{S_phi}, the correlators \ref{hight_corr} and \ref{phase_corr} have, at large $|x|$ and/or $|\tau|$, the same universal functional form if one makes the substitution
\be
n_s \, \leftrightarrow \, \chi 
\label{swap}
\ee
after subtracting non-universal additive constants 
$K_\infty  \equiv  K(\infty, \infty)   =  \langle   h^2(0,0)  \rangle $ 
and  $F_\infty  \equiv F(\infty, \infty)  = \langle \phi^2(0,0) \rangle $.
Since TQF features long-range order in both $h$ and $\phi$ fields these non-zero constants are defined for the ground state in the thermodynamic limit. This establishes the quantitatively precise
correspondence between the phase and shape fluctuations
\be
\chi [K(x,\tau) - K_\infty ] = 
n_S  [F(x,\tau) - F_\infty ] = C(x,\tau) = -  \int_{k,\omega} \,
e^{ik x+ i\omega \tau} \, \frac{k^2}{(\omega/D)^2 + k^4} \, ,
\label{Q}
\ee
with $\int_{k,\omega}\;\equiv \int d\omega dk/(2\pi)^2$. Straightforward integration over $\omega$ results in
a Gaussian integral over $k$ and the final answer 
\be
C(x,\tau)  \,=\, - {\sqrt{D} \, e^{-{x^2 \over 4 D |\tau |}} \over 4 \sqrt{\pi |\tau |}} \, \qquad \qquad (T=0,\quad  \mbox{thermodynamic limit} )\,.
\label{C_macro}
\ee

Up to the scale-invariant prefactor $\propto |\tau |^{-1/2}$, the dependence on $x$ and $\tau$ is self-similar and reduces to a dimensionless function of $x^2 /D |\tau |$. Exactly the same type of scaling takes place for iTQF states discussed in the next section \cite{Kuklov2024b}. Surprisingly, equal time correlations 
in TQF are simply absent in the TQF ground state, 
$F (x,\tau=0)\equiv 0$ (up to a non-universal short-distance  spatial decay), which is a specific example of zero-point fluctuations above the UV limit summing up to zero exactly.

Since measurements of equal time fluctuations are most natural
for a number of experimental techniques, {\it e.g.}, in the field of ultra-cold-atoms, it is important to reveal the structure of universal
space-time fluctuations in finite-size systems at non-zero temperature. The corresponding expressions follow from Equation~\ref{Q} after replacing frequency and momentum integrals with discrete sums \cite{universalTQF}. We omit here all technical details,
which are similar to what was done for iTQF in Reference~\cite{Kuklov2024b}, and proceed with presenting direct
comparison between the theoretical predictions and 
results of quantum Monte Carlo simulations for two TQF models discussed in the previous section. 

\begin{figure}[htbp]
\centering
\includegraphics[width=0.65 \columnwidth]{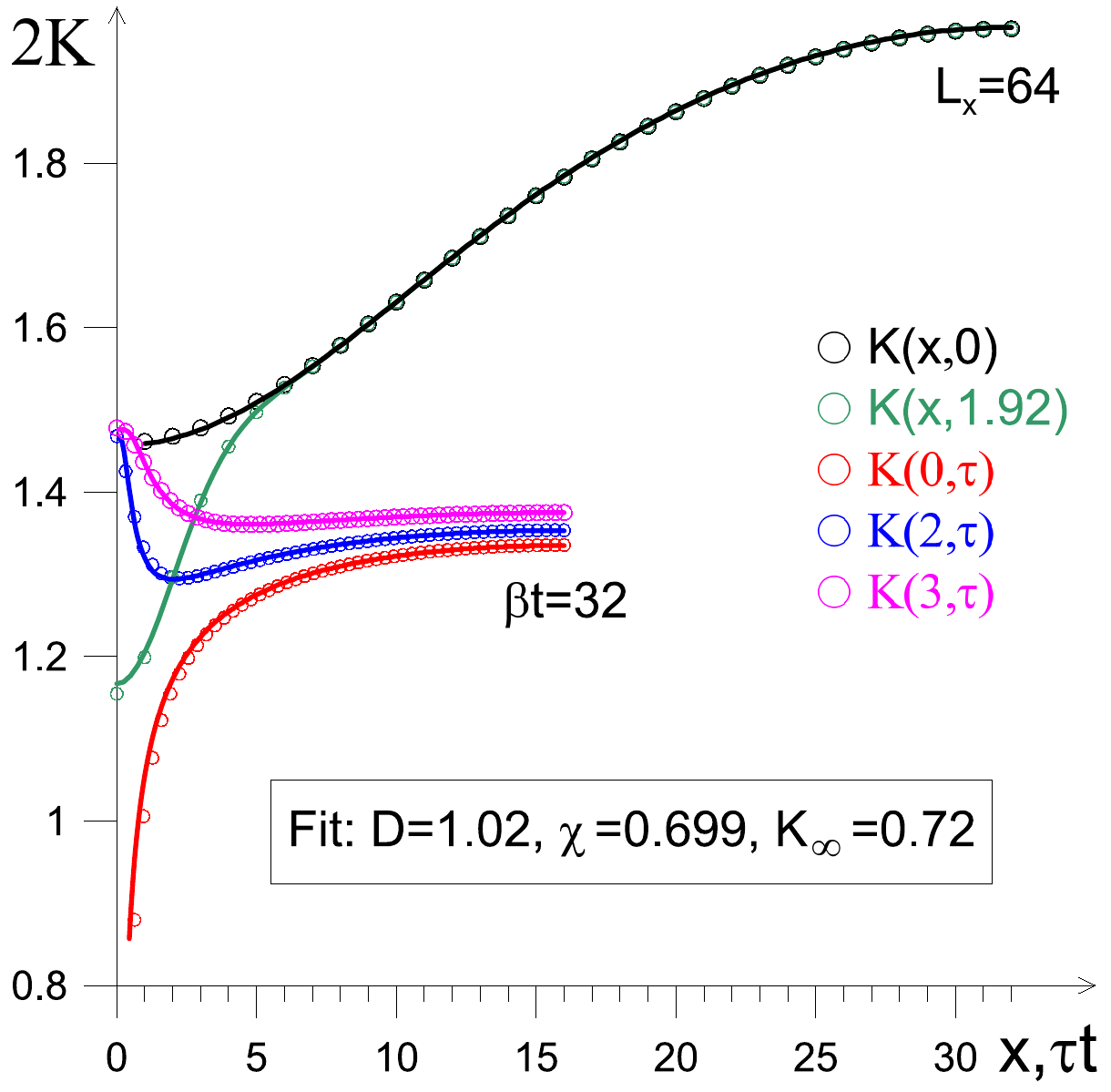}
\caption{Edge fluctuations in space and imaginary time for hard-core 
bosonic system at $V/t=-2.2$, $\beta t =32$, and system size $(L_y=32,\, L_x=64)$. The data were fitted to the TQF predictions for $2K(x,\tau)$ (solid lines). Figure adapted from Reference
\cite{universalTQF} with permission.
}
\label{K2V22}
\end{figure}

The hard-core bosonic model (isomorphic to an easy-axis ferromagnet) has been simulated using the quantum Monte Carlo Worm Algorithm  (WA) \cite{Worm}. For $V/t=-2.2$ 
the domain wall width obtained through density profiles
remains relatively large, $d/a \approx 3.15$ [it diverges 
at the SU(2) symmetric critical point $V_c/t=-2$], ensuring that
the Peierls potential for a system with $L_x=64$ sites in the $x$-direction can be neglected.  
Simulation data for $\langle [ h(x,\tau) - h(0,0) ]^2  \rangle$
are presented in {\bf Figure~\ref{K2V22}} along with their fit using the equation  $K(x,\tau) = K_\infty+C(x,\tau)/ \chi $.
Out of three fitting parameters, $K_\infty$ is responsible for the overall vertical shift and $\chi$ for the overall scale, while $D$ alone is controlling the
shapes and relative positions of all curves in {\bf Figure~\ref{K2V22}}.
The quality of agreement between the theory and simulations  of shape fluctuations demonstrated by {\bf Figure~\ref{K2V22}} is 
exceptional and leaves no doubt that we are dealing with the TQF system: while only large $x^2+\tau^2 \ge 16$ data points
were fitted, the agreement extends all the way to 
$x=1$ at $\tau=0$ and $\tau \sim 0.5$ at $x=0$.  

The ultimate confirmation comes from agreement between the simulated superfluid stiffness, $n_s = 1.492(2)$, and the result $n_s=1.49(4)$ deduced from the $n_s=D^2/\chi$ relation. This is the only situation that we know of when 
measurements of density snapshots allow one to deduce $n_s$. Similar results were obtained for the TQF setup in the 
phase-separated state of the two-component system described in previous section \cite{universalTQF}.

\subsection{Incoherent transverse quantum fluids}

\begin{figure}[htbp]
\centering
\includegraphics[width=1.0 \columnwidth]{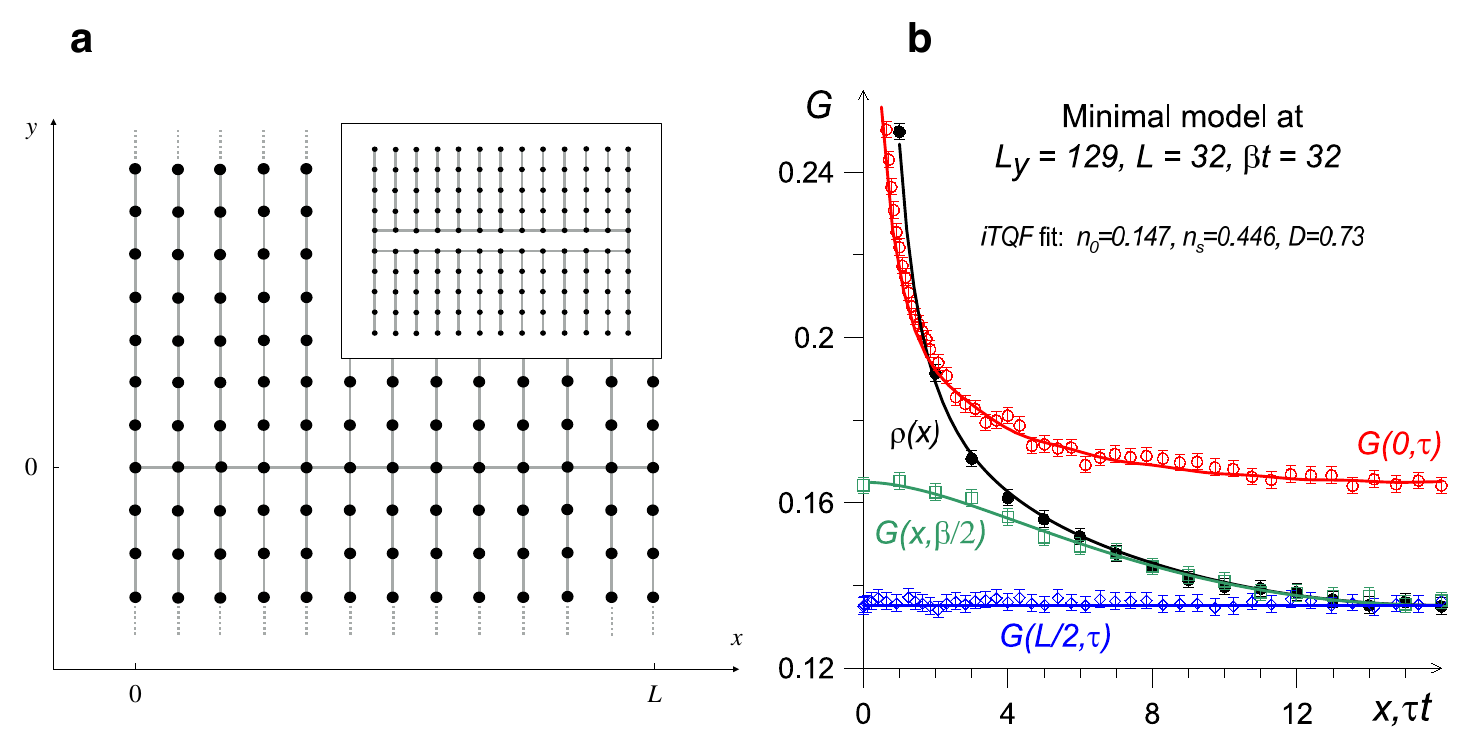}
\caption{\label{fig_itqf}
Incoherent transverse quantum fluid. (a) Minimal iTQF model: Hard-core bosons at filling factor $n = 0.5$ on the square lattice, with all hopping amplitudes between sites in the $x$ direction suppressed, except along the single row at $y = 0$. The nonzero hopping amplitudes are marked by solid lines. Inset: Image of a system of finite length with periodic boundary conditions, supporting a superflow along the central loop, which is expected to be implementable experimentally.
(b)~Single-particle Green's function $G(x,\tau)$ of the minimal model at $L=32$ and $L_y=139$ (the system sizes in the $x$ and $y$ directions, respectively) and $\beta=32/t$, with $t$ the hopping amplitude. 
Solid lines are fits to the predictions of the effective iTQF theory using
$n_0$ and $D$ as fitting parameters ($n_s=0.446$ is computed from statistics of winding number fluctuations). Figures are adapted with permission from Reference~\cite{Kuklov2024a}.
}
\end{figure}

Incoherent transverse quantum fluids (iTQF) share with TQFs the existence of true ODLRO order (Bose-Einstein condensation) in the ground state, as well as the strong topological protection of supercurrent states due to the tight-binding of instantons. The iTQFs are however a different phase of matter: they lack well-defined quasi-particles and show instead diffusive dynamics of the field of superfluid velocity. In the simplest setup, every site on a one-dimensional ring, forming the system of interest, is coupled to a gapless, one-dimensional reservoir (see {\bf Figure~\ref{fig_itqf}a}). The coupling between the system and the baths is via the phase of the particles, through which an infinite compressibility of the system particles arises, which in turn is responsible for such unusual properties in superfluids as the absence of superfluid modes and a Landau critical velocity.

Following Reference~\cite{Kuklov2024b}, the iTQF Euclidean low-energy effective action is given by
\begin{equation}
S_{\rm iTQF} = { 1 \over 2}\sum_{\omega, k}  \left[K_b|\omega|
  + n_s k^2 \right]|\phi_{\omega,k}|^2 \, ,
\label{iQTF_action}
\end{equation}
where $n_s$ is the superfluid stiffness of the system, and $K_b = (2/\pi) K$ is proportional to the Luttinger parameter $K$ of the bath. Writing the boson field as $\psi \sim e^{i\phi}$, then the disperion of the superfluid phase, $\phi(x,\tau)$, is given by $\omega = -  i Dk^2$, where $D=n_s/K_b$ is the diffusion coefficient.

For a free theory such as Equation~\ref{iQTF_action}, standard techniques can be applied to compute any correlation function. For the single-particle Green function for instance, it is possible to absorb all divergent UV contributions into a single quantity, the condensate density, which furthermore is determined by the asymptotic behavior of the Green function.
In particular, the Green function defined as
$G(x,\tau) = \langle \psi^*(x,\tau) \psi(0,0)\rangle$,
can be cast into the form \cite{Kuklov2024b}
\begin{equation} G(x,\tau) \, =\,   n_{\rm uv} \,  e^{-{1\over 2}\,  \langle \, [\, \phi(x,\tau)-\phi(0,0)\, ]^2 \, \rangle }  \, \rightarrow \, n_0 \,   e^{C(x,\tau)}  \quad \quad {\rm for} \, \, \,x^2 + D  \tau  \gg 1,
\label{G}
\end{equation}
\begin{equation}
C(x,\tau) \, =\, {D \over n_s} \int_{k,\omega}
 \,  \,  {e^{ik x+ i\omega \tau}\over |\omega| +D k^2} \, \qquad \qquad (T=0,\quad  \mbox{thermodynamic limit} )\, .
\label{Qiqtf}
\end{equation}
Qualitatively, the self-similar pattern of space--(imaginary-)time phase-phase correlations in iTQF is very reminiscent of that in TQF. While now the integration cannot be performed in terms of elementary functions, the self-similarity of the function $C(x,\tau)$ is readily seen from its integral representation:
\[
C(x,\tau) = {D  \over n_s |x|} g(\sigma) = {\sqrt{D}  \over n_s \sqrt{|\tau |} } f(\zeta=1/\sqrt{\sigma})\, , \qquad \sigma = D \vert\tau\vert /x^2\, ,
\]
\[
f(\zeta) \, \equiv \, g(1/\zeta^2)/\zeta\, , \qquad g(\sigma) \, = \,  \int_{\omega,k}  \,  {e^{ik+i\omega \sigma} \over |\omega| +  k^2}\,  = {\rm Re}
\left[ \frac{(1+i) e^{\frac{i}{4\sigma}}} {2 \sqrt{2\pi \sigma}} {\rm erfc} \left( \frac{1+i}{2 \sqrt{2 \sigma}} \right) \right]. \] 

Here we also see that the single-particle Green function approaches a constant with an asymptotic correction $\sim 1/x$ along the spatial direction and $\sim 1/\sqrt{\tau}$ in the temporal direction, in remarkable contrast to LL. 

In Reference~\cite{Kuklov2024b} closed expressions for the Green function at finite temperature and finite system size are also given and compared with quantum Monte Carlo simulations. The numeric data demonstrates excellent agreement with the effective theory (see {\bf Figure~\ref{fig_itqf}b}).

\subsection{Irrelevance of Landau criterion}

In the early days of superfluidity, Landau's argument was perceived as a necessary and sufficient condition for the system to be superfluid, and, correspondingly, was referred to as the ``Landau criterion (LC) of superfluidity." Later it was understood that superfluidity is a manifestation of the (topological) ODLRO, while LC is rather a stability criterion for uniform superflow, the violation of which does not necessarily mean that the flow will decay [see, e.g., Reference~\cite{book} for a detailed discussion.] TQF systems 
provide an excellent context to discuss the physical origin and relevance of LC \cite{Suppl}. 

In its original formulation, LC relies on the system being Galilean invariant. Since TQF is not Galilean invariant, one first needs to generalize LC to systems with asymptotic---in the long-wavelength limit---translation invariance in the absence of Galilean invariance. It is sufficient to perform the corresponding analysis at the classical-field level for bilinearized hydrodynamic Hamiltonian, $H_{\rm hd}[\eta, \phi]$, written in terms of canonically conjugate fields $\eta$ and $\phi$. For LL, these are small deviations of the coarse-grained matter density from its equilibrium value $n_0$ and 
superfluid phase from $\phi_0({\bf r}) = {\bf v}_0 \cdot {\bf r}$, respectively (in our units the bare particle mass $m=1$).
The dependence of ground-state energy density, ${\cal E}(n, v_s)$, on density and superfluid velocity magnitude, $v_s$, obeys the following hydrodynamic relations~\cite{book}:
\begin{equation}
 {\partial {\cal E}(n, v_s) \over \partial n }  \, =\, \mu\, , \qquad \quad {\partial {\cal E}(n, v_s) \over \partial v_s } \, =\,m j \, , \quad
 \label{Hydro_relations}
\end{equation}
where $j=n_sv_s$ is the supercurrent density magnitude.  

Performing an expansion of $\cal E$ around $n=n_0, v_s=v_0$, the Hamiltonian density ${\cal H}_{\rm hd}[\eta, \phi]$  to lowest-order in $(\nabla \phi, \eta)$ is given by
\begin{equation}
{\cal H}_{\rm hd}(\eta, \phi) \, =\,   {\tilde{n}_s \over 2m } \, (\nabla \phi)^2 \, +\, {\eta^2 \over 2\kappa} \,
\,+\, \lambda_s \eta \, {\bf v}_0 \cdot \nabla \phi \,  ,
\label{H_hydro}
\end{equation}
where
\begin{equation}
\tilde{n}_s\, =\,  {1\over m}\frac{\partial^2\cal E}{\partial v_0^2}\, =\, \frac{\partial j}{\partial v_0}\, =\, n_s  +v_0 \frac{\partial n_s(n_0,v_0)}{\partial v_0}\, ,
\label{tilde_n_s}
\end{equation}
and
\begin{equation}
\kappa \, =\,  {\partial n(\mu, v_0) \over \partial \mu}\, , \qquad \quad   \lambda_s \, =\, {\partial n_s(n_0,v_0)  \over \partial n_0 } \, ,
 \label{partials}
\end{equation}
where $\kappa$ is the compressibility. Note that in the absence of Galilean invariance $n_s$ can be a nontrivial function of $(n, v_s)$.

The dispersion of normal modes governed by the Hamiltonian (\ref{H_hydro}) is
\begin{equation}
\omega({\bf k}) \, =\, v_* k \, +\, \lambda_s {\bf v}_0\cdot {\bf k}\, , \qquad \quad v_*\, =\, \sqrt{\tilde{n}_s \over \kappa} \, . \quad
 \label{disp}
\end{equation}
The critical velocity, $v_0^{(c)}$,  deduced from  
the $\omega({\bf k}=0) $ condition reads
\begin{equation}
v_0^{(c)} \, =\, v_*/\lambda_s\, ,
 \label{Landau}
\end{equation}
implying that $\omega({\bf k})$ goes negative 
for some momenta when $v_0 > v_0^{(c)}$.
In the Galilean invariant superfluid, $\tilde{n}_s=n_s = n$, $\lambda_s =1$, and the critical velocity coincides with the sound velocity: $v_0^{(c)}=v_*$. 
In a general case, $\lambda_s$ can change not only its value but even the sign (like, for example, in a system of hard-core lattice bosons close to unit filling). Nevertheless, having $\lambda_s\equiv 0$ over a finite range of parameters rather than at a special fine-tuned point requires special conditions.

In TQF, the compressibility is infinite, $\kappa =\infty$, density is equivalent to the edge displacement, and 
potential-energy density may depend only on the shape deformation $\chi (\partial_x h)^2/2$. Furthermore, due to the translation and rotation invariance of the edge location and orientation $n_s$ can only depend on the curvature $\sim \partial^2_{xx} h$ of the line \cite{Kuklov2022}. Accordingly, the spectrum $\omega \sim k^2$ at finite $v_0$ acquires the term $\propto v_0 k^3$. In the $k\to 0$ limit,  this correction becomes irrelevant (equivalently, one can consider it as $\lambda_s\equiv 0$ statement), and there is no room for the Landau-type hydrodynamic instability.

\subsection{Instantons and critical current}\label{inst}

At sufficiently low temperature, thermally activated  (i.e., classical) phase slips are exponentially rare and the leading mechanism of the superflow decay is based on quantum phase 
slips---instantons in the phase field. In the exponential approximation \cite{ColemanInstantons}, the probability of the instanton nucleation,  ${\cal P}$,  is given by 
\be
{\cal P} \, \sim \, e^{-S_{\rm inst}^*} \, ,
\label{inst_exp}
\ee
where $S_{\rm inst}^*$ is an extremal value of the instanton action $S_{\rm inst}$. The problem of calculating $S_{\rm inst}^*$ is conveniently solved in the $(1+1)$-dimensional Euclidean space-time, where it has  close similarity with the problem of the vortex–anti-vortex
pair in a 2D superfluid:  The superflow generates a  “force” pulling the pair apart along the transverse (imaginary-time) direction. This force competes with the force of the vortex–anti-vortex attraction, and the extremal value $S_{\rm inst}^*$ corresponds to the vortex–anti-vortex configuration when the two forces exactly compensate each other. In order for this to happen, the vortex and anti-vortex have to share the same spatial coordinate, because the ``pulling apart" force acts in the imaginary-time direction. This leaves one with the problem of finding the maximum of the action 
\be
S_{\rm inst}(\tau) \, =\,  S_{\rm inst}^{(0)}(\tau)  -2\pi n_s m v_0 \tau   \, .
\label{S_inst}
\ee
Here $S_{\rm inst}^{(0)}(\tau)$ is the attractive potential between the vortex and anti-vortex sharing the same spatial coordinate as a function of their imaginary-time separation $\tau$. The second term in the r.h.s. is the repulsive potential generated by the superflow . This term has exactly the same form for both LL and TQF. 

The crucial difference between LL and TQF cases comes from the form of $S_{\rm inst}^{(0)}(\tau)$. In the LL case, $S_{\rm inst}^{(0)}(\tau) \propto \ln \tau$, resulting in the power-law dependence ${\cal P} \sim (v_0)^g$, with $g>0$. In both TQF and iTQF, we have $S_{\rm inst}^{(0)}(\tau) \propto \sqrt{\tau}$ \cite{Radzihovsky2023,Kuklov2024a}, which yields 
\be
{\cal P} \sim  e^{-v_c/v_0} \, .
\label{P_exp}
\ee
The parameter $v_c$ has the meaning of the characteristic velocity of the superflow below which the probability of quantum phase slip becomes exponentially small.

Specifically, in the TQF case, we have  \cite{Radzihovsky2023} 
\be
S_{\rm inst}^{(0)}(\tau)\, =\,  \frac{2\pi^{3/2}n_s}{\sqrt{D}} \sqrt{\tau} \, .
\label{V_inst}
\ee
The extremal value of $S_{\rm inst}(\tau)$ is thus reached at
$\tau=\tau_* =  \pi /(4 D v_0^2) $, and the corresponding extremal action is given by
\be 
S_{\rm inst}^* \, \equiv \, S_{\rm
  inst}(\tau_*) \, = \, \frac{ \pi^2 n_s }{2 D v_0}\, .
\label{instantonPairOpt}
\ee
Substituting this into (\ref{inst_exp}), we get (\ref{P_exp}) with
\begin{equation}
 v_c = \frac{\pi^2 n_s }{2 D}= \frac{ \pi^2 \sqrt{n_s/\chi}}{2} \, .
 \label{v_c}
 \ee

In conclusion, note that while the form of iTQF action, Equation~\ref{iQTF_action}, is qualitatively different from its TQF counterpart, Equation~\ref{S_phi}---resulting in dramatically different dynamics of the superfluid velocity field perturbations---the instanton physics in both cases is qualitatively similar \cite{Kuklov2024a}.

\section{TEMPERATURE DEPENDENCE OF CRITICAL FLUX IN HE-4: SOLVING THE PUZZLE}
 \label{sec:4}
 
 As discussed in the introduction, the STS effect  \cite{Hallock2008,Hallock2009} demonstrates three puzzling features inconsistent with any known mechanism. While the syringe effect has been readily explained in terms of the dislocation superclimb \cite{sclimb}, it took more than a decade to explain two other features: a near-exponential suppression of the flux with temperature and the sub-Ohmic I-V curve with the temperature-dependent exponent shown in {\bf Figures \ref{fig:anomalous_critical_flux}a} and {\bf \ref{fig:anomalous_critical_flux}b}, respectively.

 The key insight came from observed exponentially strong suppression of the flux by pressure in the experiment \cite{Moses} and {\it ab initio} simulations \cite{Kuklov2022}. The proposed scenario is that large temperature-induced edge dislocation shape fluctuations 
 create stress acting on the dislocation. Stressed 
 dislocation segments work as rare superflow bottlenecks (weak links where the phase slips take place) through exponential dependence of local $n_s$ on the core curvature $\sim \partial^2_x h$, see Equation~\ref{H}. Within this framework, it became possible to explain the near exponential temperature dependence of the flux ({\bf Figure~\ref{fig:anomalous_critical_flux}a}). Further analysis of the phase slips in Reference~\cite{Radzihovsky2023} provided quantitative explanation of observed I-V curves ({\bf Figure~\ref{fig:anomalous_critical_flux}b}). Since TQF features 
 a finite critical velocity $v_c$ (see Section \ref{inst}), the decay of current has nothing to do with LC for instability of superflows when the excitation spectrum is parabolic. It is also in a stark contrast with the LL case where the phase slip probability dependence on $1/v_0$ is a power-law.

\section{CONCLUSIONS AND OUTLOOK}

The field of supersolidity has taken a breathtaking journey since the first experimental observation of an abrupt drop in the period of torsional oscillator filled with solid \he4 by Kim and Chan in 2004~\cite{Kim_Chan_2004}. Current consensus is that torsion oscillator results originated from elastic, not superfluid, 
phenomena. Nevertheless, the discovery stimulated a number of research directions with  unforeseen outcomes.  
Among the earliest results, we mention (i) rigorous theoretical understanding of the nature of supersolid state of matter, with no qualitative alternatives to the incommensurate compressible state; and (ii) development of efficient continuous-space Worm algorithm that promptly led to {\it ab initio} predictions of defect-induced supersolidity in imperfect \he4 crystals. This was followed by (iii) experimental discovery of the giant quantum plasticity; and (iv) UMass Sandwich detection of superflow through imperfect \he4 crystals. The most important discovery in the context of this 
review was (v) the giant isochoric compressibility (a.k.a. syringe effect). 
Finding a framework that unifies all these observations has been an elusive task for nearly 15 years. Recently, a novel paradigm in quasi-1D superfluidity, named the transverse quantum fluid, was introduced, which not only offers an explanation for the unusual flow properties of \he4 in the UMass Sandwich setup, but also covers a broad class of other edge states in cold atomic systems, XY-magnets, and surface steps.

The adjective ``transverse" refers to the fact that an edge of this class, while behaving as 1D superfluid channel (in $x$-direction, to be specific) exchanges particles with a ``transverse" (in $y$-direction  or $y$- and $z$-directions) reservoir---with no long-range off-diagonal correlations in the $x$-direction. The role of this reservoir is to support an infinite (in the 1D sense) edge compressibility, which in its turn leads to other unusual properties, such as a quadratic dispersion (or even absence) of normal modes, irrelevance of the Landau criterion, ODLRO at $T = 0$, and the exponential dependence of the phase slip probability on the inverse flow velocity.

Superclimbing modes with parabolic dispersion are a hallmark of TQFs. Signature correlations caused by zero-point/low-temperature quantum fluctuations of these modes are readily seen in numeric simulations. They unambiguously reveal the defining feature of the superclimbing modes---canonical conjugation of the edge displacement field to the field of superfluid phase, with an implication that the superfluid stiffness can be measured from density snapshots. 

In the long-wavelength limit (justifying the harmonic approximation), the canonically-conjugate fields of the edge displacement and superfluid phase behave similarly, up to the choice of units. At zero temperature both fields are long-range ordered. This, in particular, implies the asymptotic relevance of Peierls barriers despite their exponential suppression at the microscopic scales. Academically speaking, this means that the presence of Peierls barriers---no matter how weak upon microscopic renormalization---should result in a crossover from TQF to an (anomalously compressible) LL regime at long enough wavelength and low enough temperature. However, geometrical jogs at sufficiently high density (in tilted dislocations) can eliminate the LL 
crossover and stabilize TQF \cite{MAX}.

The power-law of instanton--anti-instanton pairs in the $(1+1)$ Euclidean representation of TQF explains the ability of the system to support finite supercurrent despite its quasi-1D character: Quantum phase slips are exponentially suppressed at small current velocities. Such strong suppression of quantum phase slips is the key circumstance allowing one to theoretically interpret the unusual universal temperature dependence of the critical flux observed in the supertransport through solid \he4 experiments.

The kinematics of TQF calls for extending the instanton theory beyond the exponential approximation.  In  the translation-invariant LL,  the decay of the supercurrent is kinematically forbidden and one needs impurities and/or commensurate external potential to absorb the momentum released when the phase winding number changes due to a quantum phase slip. In TQF, the decay of the supercurrent into elementary excitations is kinematically allowed due to the quadratic dispersion. However, at small values of the flow velocity this involves a large number of elementary excitations \cite{Radzihovsky2023}. It is thus important to understand whether the multi-excitation decay can compete with the impurity-assisted phase slips.

Putting superclimbing TQF systems in a broader context,  it is worth noting that the iTQF system shown in {\bf Figure~\ref{fig_itqf}} is not the only one studied in the past: more examples are discussed in Reference \cite{Kuklov2024a}.

\section{SUMMARY POINTS}

\begin{enumerate}

\item The effect of supertransport through imperfect \he4 crystal is accompanied by an even more striking phenomenon of giant isochoric compressibility---the so-called syringe effect. A natural, numerically verified (and the only existing) interpretation of the intrinsic connection between the two effects is in terms of the superclimb of edge dislocations.\vspace{1mm} 
\item An effective field-theoretical description of superclimbing edge dislocation reveals a new paradigm of quasi/pseudo-1D superfluids---the so-called transverse quantum fluid (TQF).\vspace{1mm}
\item It has been suggested theoretically, and demonstrated by {\it ab initio} simulations, that the TQF state takes place in a variety of setups, such as the superfluid edge of a self-bound droplet of hard-core bosons on a two-dimensional (2D) lattice, a Bloch domain wall in an easy-axis ferromagnet, and a phase separated state of two-component bosonic Mott insulators with the boundary in the counter-superfluid phase (or in a phase of two-component superfluid) on a 2D lattice.\vspace{1mm}
\item The unifying feature of the TQF systems---responsible for their unusual superfluid properties---is the infinite effective (one-dimensional) compressibility emerging as a result of particle exchange with  a {\it nonsuperfluid} particle reservoir in the transverse direction.\vspace{1mm}
\item From the conceptual point of view, the TQF state is a striking demonstration of the conditional character of many dogmas associated  with superfluidity, including the necessity of elementary excitations, in general, and the ones obeying the Landau criterion in particular.\vspace{1mm}
\item Superclimbing modes  with parabolic dispersion are a hallmark of TQFs. Signature correlations caused by zero-point/low-temperature quantum fluctuations of these modes reveal the defining feature of the superclimbing modes---canonical conjugation of the edge displacement field to the superfluid phase field.\vspace{1mm}
\item In the long-wavelength limit, the canonically-conjugate fields of the edge displacement and superfluid phase behave similarly, up to the choice of units. At zero temperature both fields are long-range ordered. This, in particular, implies the asymptotic relevance of Peierls barriers despite their exponential suppression at the microscopic scales, resulting in a crossover from TQF to an (anomalously compressible) LL regime at long enough wavelength and low enough temperature.\vspace{1mm}
\item Tight-binding of instanton--anti-instanton pairs in the $(1+1)$ Euclidean representation of TQF explains the ability of the system to support finite supercurrent despite its quasi-1D character: Quantum phase slips are exponentially suppressed at low values of the supercurrent, which is the key circumstance allowing one to theoretically interpret the universal and most unusual temperature dependence of the critical flux observed in the experiment on the supertransport through solid \he4.
\end{enumerate}

\section{FUTURE ISSUES}
\begin{enumerate}
\item The connection between quantum plasticity and superclimb is an open outstanding topic. \vspace{1mm}
\item While barely mentioning the screw dislocation in this review, it should be recognized that this dislocation can transform--- upon bias---into a spiral with edge segments which can also perform superclimb. This effect deserves a detailed study.\vspace{1mm}
\item Implementation of TQFs with ultracold atoms and magnets is of prime fundamental interest.\vspace{1mm}
\item Signature statistical and dynamical properties of superclimbing modes---specific correlations of the hight fluctuations and the quadratic dispersion law---are yet to be observed experimentally in any system.\vspace{1mm}
\item Of special interest would be an experimental study of the crossover from TQF to LL regimes---in various experimental settings and in terms of both statistical and dynamical properties.\vspace{1mm}
\item The outstanding theoretical challenge is to develop a theory of quantum phase slips beyond exponential approximation.\vspace{1mm}
\item Identifying new members of the broadly understood TQF family is an interesting direction for future research.

\end{enumerate}

\section{ACKNOWLEDGMENTS}
We thank Massimo Boninsegni and Leo Radzihovsky---our main collaborators on projects that led to this review---for fruitful discussions and useful comments. 
A.K., B.S., and N.P. acknowledge support from the National Science Foundation under Grants DMR-2335905 and DMR-2335904, and L.P. from the Deutsche Forschungsgemeinschaft (DFG, German Research Foundation) under Germany's Excellence Strategy -- EXC -- 2111 - 390814868.

\end{document}